\documentclass[usenatbib,onecolumn]{mn2e}
\usepackage{amsfonts}
\usepackage{amsmath}
\usepackage{graphicx}
\usepackage{natbib}

\def\apjl{Astrophys.\ J.\ Lett.}
\def\mnras{Mon.\ Not.\ R.\ Astron.\ Soc.}

\def\aap{Astron.\ Astrophys.}
\def\apj{Astrophys.\ J.}

\def\prd{Phys.\ Rev.\ D}

\title[Dark matter PDF in the $f_{nl}$ model]
      {The nonlinear probability distribution function in models with 
       local primordial non-Gaussianity}

\author[T. Y. Lam \& R. K. Sheth]
 {Tsz Yan Lam\thanks{E-mail:  tylam@sas.upenn.edu, shethrk@physics.upenn.edu}
  \& Ravi K. Sheth\footnotemark[1]\\
 Department of Physics \& Astronomy, University of Pennsylvania, 
 209 S. 33rd Street, Philadelphia, PA 19104, USA}

\newcommand{\bm}[1]{{\mbox{\boldmath $#1$}}}

\begin{document}
\pagerange{\pageref{firstpage}--\pageref{lastpage}}

\maketitle

\label{firstpage}

\begin{abstract}
We use the spherical evolution approximation to investigate 
nonlinear evolution from the non-Gaussian initial conditions 
characteristic of the local $f_{nl}$ model.  
We provide an analytic formula for the nonlinearly evolved 
probability distribution function of the dark matter which shows 
that the underdense tail of the nonlinear PDF in the $f_{nl}$ model 
should differ significantly from that for Gaussian initial conditions.  
Measurements of the underdense tail in numerical simulations may 
be affected by discreteness effects, and we use a Poisson counting 
model to describe this effect.  
Once this has been accounted for, our model is in good quantitative 
agreement with the simulations.
\end{abstract}


\section{Introduction}
The most common inflation model (the single scalar field, slow-roll 
inflation) predicts the primordial perturbations to be approximately 
described by the Gaussian statistics. Detections of non-gaussianity 
can discriminate between different inflation models, so the study of 
various cosmological probes of primordial non-gaussianity has 
attracted much recent attention.  

For example, the CMB has been used to constrain non-gaussianity 
\citep{cmb5yr,hikageetal08,yw08,mhlm08}.  
In addition, various aspects of large scale structure have also 
been examined as probes of non-gaussianity.  These include 
halo abundances and bias 
\citep{lmv88,rb00,mvj00,fnlverde,dalaletal08,mv08,cvm08,at08,
shshp08,fnlvincent}; higher order effects on the galaxy power spectrum 
\citep{mcdonald08,tkm08}; and the galaxy bispectrum \citep{ssz04,sk07}.  
And, \citet{slosar08} has discussed how to optimize the 
constraint on $f_{nl}$ by applying weightings on subsamples from 
a single tracer.

Recently, \citet{grossi08} used N-body simulations to study the effect of 
primordial non-gaussianity on the nonlinear probability distribution 
function (PDF) of the dark matter field.  They found that departures 
from the Gaussian prediction are strongest in underdense regions 
and on small scales.  The main goal of the present study is to 
provide some analytic understanding of their results.  We do this 
by studying the nonlinear evolution of the dark matter PDF.  
When the initial conditions are Gaussian, then the nonlinear 
evolution can be modelled by the spherical and ellipsoidal collapse 
models \citep{lamshethreal,lamshethred}. 
In these models, the evolution of the PDF depends on two ingredients:  
a model for the dynamical evolution from an initial state to a final 
one, and the appropriate average over the initial states.  
In particular, the spherical and ellipsoidal collapse models 
approximate the nonlinear dynamics as 1$\to$1 or 3$\to$1 mappings 
from the initial to the final state that do not depend on the 
Gaussianity of the initial conditions.  Primordial non-Gaussianity 
enters only because it determines the initial set of states.  
As a result, essentially all of the machinery developed for the 
Gaussian case can be simply carried over to the present study.  

We recap the definitions of the local non-gaussian $f_{nl}$ model 
in Section~\ref{section:nosmooth}.  
The effect of smoothing the initial distribution is described in 
Section~\ref{section:initC}. 
Section~\ref{section:nlpdf} describes the nonlinear PDF associated 
with the spherical evolution model and compares the results with the 
Gaussian case.  Section~\ref{section:sims} compares our predictions 
with measurements in simulations.  
We summarise our results in Section~\ref{section:discussion}.


\section{The PDF of the density in the local non-gaussian model}
\label{section:def}
In the $f_{nl}$ models of current interest, the initial density 
fluctuation field is only mildly perturbed from the Gaussian.  
Therefore, following \cite{fnlverde} and \cite{fnlvincent}, we use 
the Edgeworth expansion as a convenient way to summarize our results.  
Because we are interested in mass scales which are substantially 
larger than that of a single collapsed halo -- the regimes studied 
by \cite{fnlverde} and \cite{fnlvincent} -- we are certainly in 
the regime where the Edgeworth expansion is a useful approximation.  

\subsection{The unsmoothed initial distribution}\label{section:nosmooth}
The primordial perturbation potential $\Phi$ of the local non-gaussian 
field is 
\begin{equation}
 \Phi = \phi + f_{nl}(\phi^2 - \langle \phi^2\rangle), \label{eqn:fnl}
\end{equation}
where $\phi$ is a Gaussian potential field and $f_{nl}$ is a scalar.
We will use $P_{\phi}(k)$ to represent the power spectrum of $\phi$; 
in what follows we will set $P_{\phi}(k)= Ak^{n_s-4}$, where 
$n_s\approx 1$, and $A$ is a normalization constant that is fixed by 
requiring that the rms fluctuations in the associated non-Gaussian 
initial density field (which we will define shortly) have value $\sigma_8$.

We define ${\bm M}$ as the real, symmetric $3\times 3$ tensor whose 
components are proportional to the second order derivatives of the 
potential $\Phi$: 
\begin{equation}
 \Phi_{ij} \equiv \phi_{ij} + 2f_{nl}(\phi_i\phi_j + \phi\phi_{ij}),
\end{equation}
where $\phi_i = \partial_i \phi$ and 
      $\phi_{ij} = \partial_i\partial_j \phi$. 
We will sometimes refer to ${\bm M}$ as the shear or deformation tensor 
associated with the potential $\Phi$.  

Correlations between the $\Phi_{ij}$ will be very useful in what 
follows.  These depend on the correlations between $\phi$ and its 
derivatives.  However, because $\phi$ is Gaussian, they can be 
computed easily.  Specifically, 
\begin{align}
 \langle\phi\phi\rangle & = \sigma_0^2 
 & \langle\phi_i\phi\rangle & = 0, 
 & \langle\phi_i\phi_j\rangle & = \frac{\sigma_1^2}{3}\delta_{ij}\\
 \langle\phi_{ij}\phi\rangle & = -\frac{\sigma_1^2}{3}\delta_{ij} 
 & \langle\phi_{ij}\phi_{k}\rangle & = 0,
 & \langle\phi_{ij}\phi_{kl}\rangle &= \frac{\sigma_2^2}{15}\delta_{ijkl} 
\end{align}
where $\delta_{ij}$ is the dirac delta function, and we have defined
\begin{align}
\delta_{ijkl}  & \equiv
 \delta_{ij}\delta_{kl} + \delta_{ik}\delta_{jl} + \delta_{il}\delta_{jk},\\
 \sigma_j^2 & \equiv \int {\rm d}{\bm k}\,P_\phi(\bm k)\,M^2(k)\,
                                          W^2(kR)\,(-k^2)^j,
 \label{eqn:sigmaj}
\end{align}
where
\begin{equation}
 M(k) \equiv \frac{3c^2}{5\Omega_m H_0^2}\,T(k)
 \label{eqn:mk}
\end{equation}
and $T(k)$ is the CDM transfer function.  (Strictly speaking, 
we are currently interested in the limit in which $W=1$.  We have 
defined the more general expression so that it can be used in the 
following subsections.)  Thus, 
\begin{align}
 \langle \Phi_{ij}\,\Phi_{kl} \rangle 
  & = \frac{\sigma_{NG}^2}{15}\delta_{ijkl},\qquad{\rm and} 
      \label{eqn:2ndorder}\\
 \langle \Phi_{ij}\,\Phi_{kl}\,\Phi_{mn} \rangle 
  & = 2f_{nl}\frac{\gamma^3_{NG}}{135} \left[\delta_{ij}\delta_{klmn} 
      + \delta_{kl}\delta_{ijmn} + \delta_{mn}\delta_{ijkl} \right] 
      + \mathcal{O}(f_{nl}^3),
\label{eqn:3rdorder}
\end{align}
where 
\begin{equation}
 \frac{\sigma^2_{NG}}{15}  = \frac{\sigma_2^2}{15} 
   + 4f_{nl}^2\left[ \left(\frac{\sigma_1^2}{3}\right)^2
   + \frac{\sigma_0^2\sigma_2^2}{15}\right] \qquad {\rm and}\qquad
 \frac{\gamma^3_{NG}}{135}  = -2\frac{\sigma_1^2}{3}\frac{\sigma_2^2}{15} .
\end{equation}


\subsection{The smoothed initial field}\label{section:initC}
In the spherical evolution model, the quantity of interest is the 
initial overdensity $\delta_l$ smoothed on some scale $V$.  
Most of the complication in $f_{nl}$ models arises from the fact 
that the effect of smoothing is non-trivial.  This nontriviality 
is a consequence of the fact that a smoothed Gaussian field is itself 
Gaussian, but this self-similarity does not hold for generic random 
fields.  

We now calculate the distribution function of the initial 
overdensity $\delta_l$ smoothed on scale $R$ in the $f_{nl}$ model. 
Our goal is to approximate this initial PDF using the 
Edgeworth expansion: 
\begin{equation}
 p(\delta_l |R)\, {\rm d}\delta_l =  \frac{e^{-\nu^2(R)/2}}{\sqrt{2\pi}}
  \left[1 + \frac{\sigma_{NG}(R) S_3(R)}{6}H_3(\nu(R)) \right]\ {\rm d}\nu(R),
\label{eqn:pdfdeltal}
\end{equation}
where $\nu(R) = \delta_l/\sigma_{NG}(R)$ and $H_3(\nu) = \nu(\nu^2 - 3)$.  
Therefore, we must specify how to calculate $\sigma_{NG}$ and 
$\sigma_{NG}\,S_3$.

The Fourier transform of the initial overdensity is related to the 
Fourier transform of $\Phi$ by $\delta_k \equiv -k^2\, M(k)\,\Phi_k$.
So spherical symmetry implies that the power spectrum and bispectrum 
of $\delta_l$ are 
\begin{align}
 P_{\delta_l}(k) & = (-k^2)^2 M^2(k)P_{\Phi}(k) 
               = k^4 M^2(k)\left[ P_{\phi}(k) + \frac{2f_{nl}^2}{(2\pi)^3} 
             \int {\rm d} {\bm q} P_{\phi}(q)P_{\phi}(|\bm k - \bm q|) \right],
   \label{eqn:pkij}\\
 B_{\delta_l}(k_1,k_2,k_{12}) & = 
      (-k_{1}^2)(-k_{2}^2)(-k_{12}^2)M(k_1)M(k_2)M(k_{12})
      B_{\Phi}(k_1,k_2,k_{12}),  \label{eqn:bkij} \\
 B_{\Phi}(k_1,k_2,k_{12}) & \equiv 
    2f_{nl}\,\left[ P_{\phi}(k_1)P_{\phi}(k_2) + 
    {\rm cyclic} \right] + \mathcal{O}(f_{nl}^3) \label{eqn:bkphi}.
\end{align}
From equations~(\ref{eqn:2ndorder}) and (\ref{eqn:3rdorder}), we have
\begin{align}
 \sigma_{NG}^2  & \equiv \langle\delta_l^2\rangle
     = \frac{1}{(2\pi)^3}\int \frac{{\rm d} k}{k} 4
       \pi k^7 M^2(k)P_{\Phi}(k) W^2(kR), \label{eqn:variance} \\
 \sigma_{NG} S_3 & \equiv 
       \frac{\langle\delta_l^3\rangle}{\langle\delta_l^2\rangle^{3/2}} 
     = \frac{2f_{nl}\gamma_{NG}^3 + \mathcal{O}(f_{nl}^3)}{\sigma_{NG}^3}
       \qquad {\rm where}\\
 2f_{nl}\gamma_{NG}^3(R) & = -\frac{2}{(2\pi)^4}
               \int  \frac{{\rm d} k_1}{k_1} k_1^3 W(k_1R)
               \int  \frac{{\rm d} k_2}{k_2} k_2^3 W(k_2R)
               \int {\rm d}\mu_{12}\, W(k_{12}R) \,
               B_{\delta_l}(k_1,k_2,k_{12}), \label{eqn:skewness}
\end{align}
with $\mu_{12} \equiv \cos\theta_{12}$, and $\theta_{12}$ is the 
angle between ${\bm k_1}$ and ${\bm k_2}$:  
$k_{12}^2 \equiv k_1^2 + k_2^2 + 2k_1k_2\,\mu_{12}$. 

To second order in $f_{nl}$, the variance $\sigma_{NG}^2(R)$ is the 
sum of two terms; one is the same as for the Gaussian, and the 
second, which is proportional to $f_{nl}^2$, involves a convolution 
of the power spectrum $P_{\phi}$ with itself.  This term has two 
infra-red singularities (at $q = 0$ and ${\bm k} = {\bm q}$ 
respectively) for $n_s < 4$.  
Fortunately, these can be removed by rewriting the power spectrum as 
\begin{equation}
 P_{\Phi}(k) = P_{\phi}(k) + \frac{2f_{nl}^2}{(2\pi)^3}
   \int {\rm d}\, {\bm q} \left[P_{\phi}(q)P_{\phi}(|{\bm k} - {\bm q}|)
  - P_{\phi}(k)P_{\phi}(q) - P_{\phi}(k)P_{\phi}(|{\bm k} - {\bm q}|)\right],
\end{equation}
where $P_{\phi}(k)$ has a new normalization \citep{mcdonald08}. 
For $|f_{nl}| \le 100$, this renormalization is just a few percent 
effect, but the procedure is essential for removing the singularities.

Equation~(\ref{eqn:skewness}) indicates that
 $\langle\delta_l^3\rangle$ scales linearly with $f_{nl}$. 
On scales larger than about $100h^{-1}$Mpc, the integral which defines 
$\gamma^3_{NG}$ can be approximated analytically \citep{ssz04}, 
but on smaller scales, the integral must be evaluated numerically.  
For models of interest, $\sigma S_3$ is only weakly scale dependent:  
e.g., it is $\approx -0.02 (f_{nl}/100)$ on $\sim 100h^{-1}$Mpc, 
and is less than a factor of two larger on scales $(\sim 1h^{-1}$Mpc) 
\cite[e.g., Figure~1 in][]{ssz04}.
%
%
%
This means that $|\sigma_{NG}\, S_3| \ll 1$ on the scales of interest 
in this paper, justifying our use of the Edgeworth expansion.  
In addition, note that if $\sigma_{NG}(R) S_3(R)$ were independent 
of $R$, then the Edgeworth expansion would be a function of $\nu$ 
only.  In this case, the initial (non-Gaussian) PDF would be 
scale-independent, in the sense that the PDF would have the same 
functional form for all smoothing scales, just as it does for the 
Gaussian.  
Over a sufficiently narrow range of scales, this is a reasonable 
approximation.


\subsection{The PDF of the smoothed, evolved, nonlinear overdensity} \label{section:nlpdf}
The previous section showed how to calculate the distribution 
of the initial overdensity $\delta_l$.
Nonlinear evolution changes this distributions; this section shows 
how to estimate the evolved, nonlinear PDF of the overdensity.  
As noted in the introduction, this can be done by following the 
same steps as in Lam \& Sheth (2008), but with the non-Gaussian 
initial distributions.  For brevity, we show results for the 
spherical evolution model only.  

In the spherical model, the nonlinear overdensity in a region of 
volume $V$ containing mass $M$ is solely determined by the linear 
overdensity $\delta_l$ through a $1\to 1$ mapping:  
\begin{equation}
 \rho \equiv 1 + \delta = \frac{M}{\bar{\rho}V} 
      = \left(1 - \frac{\delta_l}{\delta_c}\right)^{-\delta_c}, 
 \label{eqn:SCapprox}
\end{equation}
\citep{b94,rks98}, where $\delta_c\approx 5/3$ and the exact 
value depends weakly on the cosmology. 
In this study we will use $\delta_c = 1.66$ 
which corresponds to the $\Lambda$CDM cosmology.
The PDF of the nonlinear overdensity in a volume $V$ associated with 
spherical collapse is 
\begin{equation}
 \rho^2 p(\rho|V) = p_{NG}\Bigl(\delta_l(\rho)|V_l(\rho)\Bigr)\, \nu\, 
                    \frac{{\rm d} \ln\nu}{{\rm d}\ln\rho},
\end{equation}
where $p_{NG}(\delta_l|V_l)$ is the initial PDF of $\delta_l$ at a given 
smoothing scale $V_l$.  This with equation~(\ref{eqn:pdfdeltal}) 
implies that 
\begin{eqnarray}
 \rho^2 p(\rho|V) &=& \frac{1}{\sqrt{2\pi\sigma^2(\rho)}}
          \exp\left[ -\frac{\delta_l^2(\rho)}{2\sigma^2(\rho)}\right]
 \left[ 1 -\frac{\delta_l(\rho)}{\delta_c} + 
 \frac{\gamma_{\sigma}}{6}\delta_l(\rho)\right]
 \left[ 1 + \frac{\sigma(\rho)S_3(\rho)}{6}\,
         H_3\left(\frac{\delta_l(\rho)}{\sigma(\rho)}\right)\right],\nonumber\\
  &=& \rho^2 p_{\rm G}(\rho|V)\,
 \left[ 1 + \frac{\sigma(\rho)S_3(\rho)}{6}\,
            H_3\left(\frac{\delta_l(\rho)}{\sigma(\rho)}\right)\right],
\label{eqn:pdfsph}
\end{eqnarray}
where $p_{\rm G}$ denotes the smoothed nonlinear PDF assoociated with 
Gaussian initial conditions, 
$\sigma(\rho)=\sigma_{NG}(R = (3M/4\pi\bar\rho)^{1/3})$, and 
$\gamma_{\sigma} \equiv -3\,{\rm d} \ln \sigma^2/{\rm d} \ln M$ 
is not to be confused with the skewness parameter we defined earlier.  
Finally, we set $\rho' \equiv N\rho$ and 
 $\rho'^2\, p'(\rho') \equiv \rho^2\, p(\rho)$ to ensure that 
$\int {\rm d}\rho'\,\rho'\,p'(\rho')$ and 
$\int {\rm d}\rho'\,p'(\rho')$ both equal unity
\citep[see discussion in][]{lamshethred,lamshethreal}.  

The non-gaussian modification contributes the final term of the 
right hand side of equation~(\ref{eqn:pdfsph}).  Since $\delta_l=0$ 
when $\rho=1$, there is little or no correction to the Gaussian 
case at $\rho\approx 1$.  However, there is an effect at the low 
and high density tails.  To see what it is, note that $\sigma S_3$ 
is only a weak function of $R$, so it is a weak function of $\rho$, 
and hence the main $\rho$ dependence is due to $H_3$.  To see what 
this dependence is, suppose that $\gamma_\sigma=-6/5$ 
(this is close to its actual value on the scales we consider when 
comparing with simulations in the next section).  Then 
$\delta_l/\sigma \approx (5/3)(1-\rho^{-3/5})/(\sigma_{\rm V}\rho^{-1/5})
                 \approx (5/3\sigma_{\rm V})(\rho^{1/5}-\rho^{-2/5})$,
where $\sigma_{\rm V}$ is the variance of the initial field when 
$\rho=1$.  When $\rho\gg 1$, then 
$\delta_l/\sigma \approx (5/3\sigma_{\rm V})\,\rho^{1/5}$, so 
$H_3\propto \rho^{3/5}$.  
As a result, for $f_{nl}>0$, the high density tail of the PDF is 
increasing suppressed compared to the Gaussian case as $\rho$ 
increases.  At low densities, $\rho\ll 1$, then 
$\delta_l/\sigma \approx -(5/3\sigma_{\rm V})\,\rho^{-2/5}$, so 
$H_3\propto -\rho^{-6/5}$:  the low density tail is enhanced 
for $f_{nl}>0$.  The dependence on $\rho$ is stronger for $\rho\ll 1$ 
than for $\rho\gg 1$, so we expect the underdense tail to be a good 
probe of $f_{nl}$.  
For sufficiently large $|\delta_l/\sigma|$, the non-Gaussian piece 
can be negative, signaling that our truncation of the Edgeworth 
expansion was inappropriate.  
Fortunately, we are generally only interested in scales that are not 
significantly affected by this truncation problem.

\begin{figure}
\centering
\includegraphics[width=0.8\linewidth]{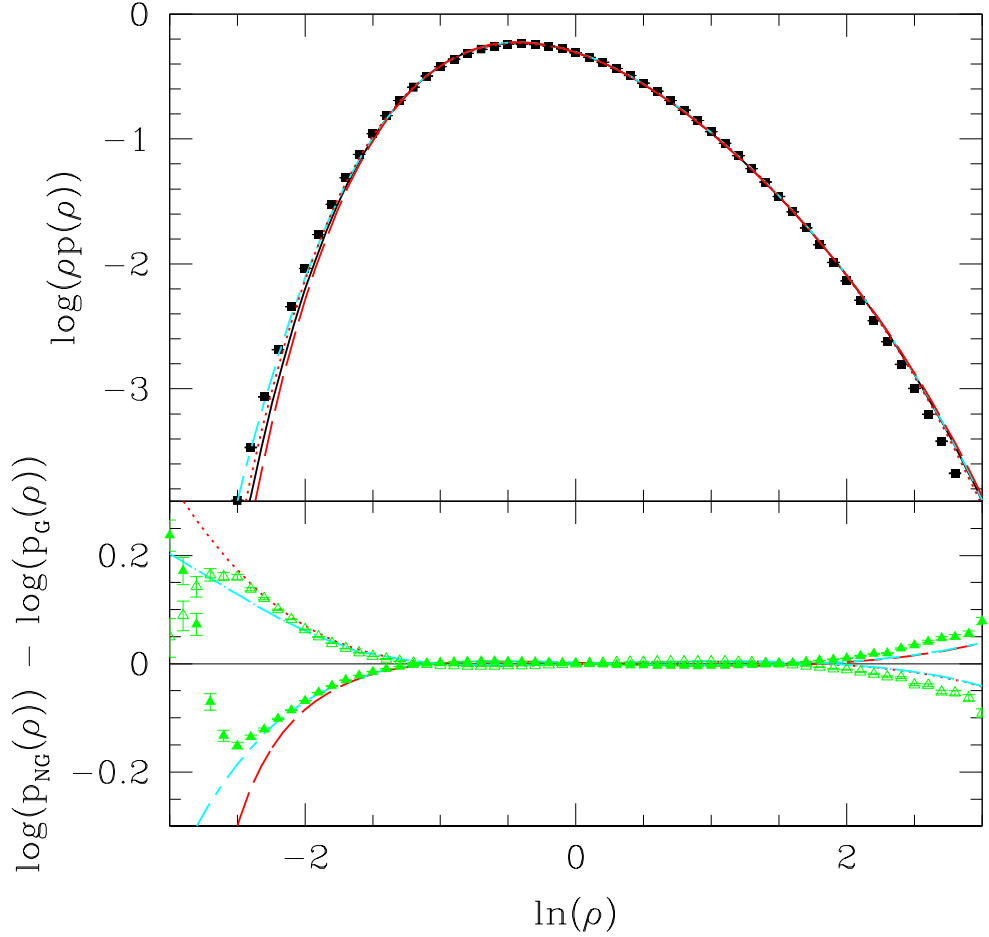}
\caption{Nonlinear overdensity PDF using the spherical 
         collapse model in cells of radius $8 h^{-1}{\rm Mpc}$.  
         Long-dashed, solid, and dotted curves in the upper panel 
         show $\ln (\rho\,p(\rho))$ (from equation~\ref{eqn:pdfsph}) 
         as a function of $\ln\rho$ for $f_{nl} = -100, 0$ and 100.  
         Symbols show the PDF measured in the $f_{nl}=100$ 
         simulation, and long-dashed-short-dashed curve shows the 
         associated Poisson-sampled prediction (equation~\ref{eqn:poisson}).
         The lower panels show the log of the ratio between the 
         $f_{nl}\ne 0$ predictions and that for Gaussian initial 
         conditions, for which $f_{nl}=0$. Filled and empty symbols 
         are similar ratios of the measured PDFs for $f_{nl}=-100$ 
         and 100 respectively. }
\label{fig:sphPTr8}
\end{figure}

\subsection{Comparison with simulations}\label{section:sims}
We compare the predictions of our approach with measurements of 
the nonlinear PDF in numerical simulations from \cite{fnlvincent}.  
These followed the evolution of $1024^3$ particles in a 
periodic cube of sides $1600\,h^{-1}$Mpc.  The background 
cosmology is spatially flat, dominated by a cosmological 
constant, having
 $(\Omega_m,\Omega_b,n_s,h,\sigma_8)=(0.279,0.0462,n_s=0.96,0.7,0.81)$, 
so the particle mass was $3\times 10^{11}h^{-1}M_\odot$.
The simulations sample the density field using discrete particles.  
This produces discreteness effects which are largely irrelevant, 
except in the least dense tails of the PDF.  
(E.g., the average number of particles in spherical cells with 
radius $4h^{-1}$Mpc is $\approx 60$.  
So discreteness effects are severe at $\rho < 1/6$.)
We account for this using the Poisson model:  
\begin{equation}
 p(N|V) = \int dM\,p(M|V)\,p(N|M)
 \label{eqn:poisson}
\end{equation}
where $p(N|M,V) = (M/m_p)^N \exp(-M/m_p)/N!$, with $m_p$ equal to 
the particle mass \citep{rks96}.  
We then set $\rho \equiv N/\bar nV$ when plotting the results.

The symbols in the top panel of Figure~\ref{fig:sphPTr8} show the 
measured PDF for counts in spheres of radius $8h^{-1}$Mpc for 
$f_{nl} = 100$.  
The dashed curve which is closest to these symbols shows 
equation~(\ref{eqn:poisson}) for $f_{nl}=100$.  
The dashed, solid and dotted curves show the predictions of 
equation~(\ref{eqn:pdfsph}) for $f_{nl}=-100$, 0 and 100.  
The bottom panel shows the ratio of the counts in the $f_{nl}$ 
models to those when $f_{nl}=0$.  The outer set of curves show 
equation~(\ref{eqn:pdfsph}) and the inner set show the effect 
of accounting for discreteness effects using equation~(\ref{eqn:poisson}).
This shows that the PDF for positive $f_{nl}$ is slightly skewed 
towards underdense regions, the reverse is true for negative $f_{nl}$, 
and our model provides an excellent description of these trends.  

\begin{figure}
\centering
\includegraphics[width=0.8\linewidth]{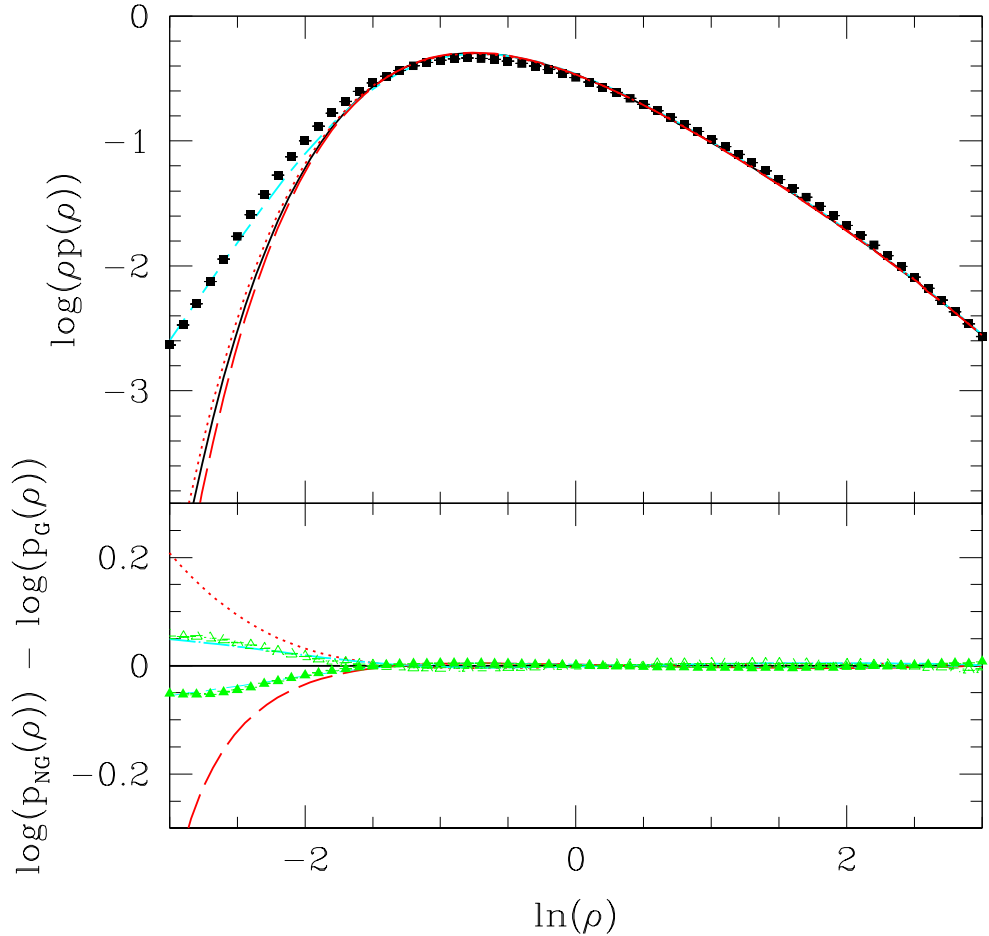}
\caption{Same as previous figure, but for cells of radius 
         $4 h^{-1}{\rm Mpc}$.}
\label{fig:sphPTr4}
\end{figure}

Figure~\ref{fig:sphPTr4} shows a similar analysis of the PDF on 
smaller scales, for which the discreteness effects are more 
pronounced: at small $\rho$, the symbols in the top panel lie 
well above the curves associated with equation~(\ref{eqn:pdfsph}).  
However, our Poisson model for the discreteness effect appears 
to be rather good:  the prediction associated with 
equation~(\ref{eqn:poisson}) provides a good description of the 
measurements.  The bottom panel shows that discreteness effects 
tend to wash out the differences between the different $f_{nl}$ 
runs, but that our Poisson model does an excellent job of 
accounting for this effect (the predictions at small $\rho$ are 
almost indistinguishable from the measurements).

\section{Discussion} \label{section:discussion}
We used the spherical evolution model to study the nonlinear evolved 
probability distribution function of the dark matter density field 
in the local primordial non-Gaussian $f_{nl}$ model.  (The spherical 
model is able to provide a good description of the nonlinear PDF 
when $f_{nl}=0$.)
For currently acceptable values of $f_{nl}$, our approach shows that 
the signatures of primordial non-Gaussianity are small, but are most 
evident in underdense regions (equation~\ref{eqn:pdfsph} and related 
discussion).
The PDF measured in simulations can be affected by discreteness 
effects, especially in small underdense cells.  The effect of 
this can be approximated by using a Poisson counting model 
(equation~\ref{eqn:poisson}).  
Once this is done, our model is in very good agreement with 
measurements in numerical simulations (Figures~\ref{fig:sphPTr8} 
and~\ref{fig:sphPTr4}), so we hope our equation~(\ref{eqn:pdfsph}) 
will be useful in studies which require knowledge of the nonlinear 
evolved PDF.  

Recently, following the same logic for why cluster abundances should 
be good probes of primordian non-Gaussianity, \cite{kvj08} have 
suggested that void abundances should also be good probes.  Our 
results provide further motivation for studying underdense regions.  
We are in the process of developing a more complete model of voids 
and void shapes \cite[following][]{sw04}.

Finally, note that the parameter which describes the non-Gaussianity 
in the smoothed initial and final fields, $\sigma\,S_3$, is only weakly 
scale-dependent.  Therefore, our analysis indicates that the 
non-gaussian distribution of the resulting nonlinear PDF 
(equation~\ref{eqn:pdfsph}) can be written in terms of the scaled 
variable $\nu=\delta_l/\sigma$.  So one could formulate a reconstruction 
of the initial $f_{nl}$ field analogously to how this is done for the 
Gaussian case \citep{lamshethreal}.  We have not pursued this further.

\section*{Acknowledgements}
We thank V. Desjacques for help with measuring the PDFs in his 
simulations and R. Scoccimarro and E. Komatsu for many helpful 
discussions.
Thanks also to S. Cole, K. Dolag, M. Grossi, W. Hu, Y. P. Jing, Z. Ma, 
T. Nishimichi, and R. Scoccimarro for discussions about resolution 
effects on the setting up of initial conditions in simulations.

\label{lastpage}
\end{document}